\documentclass[amssymb,aps,eqsecnum]{revtex4}
\usepackage{amsmath}
\usepackage{enumerate}
\usepackage{graphicx}
\DeclareGraphicsExtensions{.pdf,.png,.jpg}
\begin{document}
\title{Electromagnetic momentum and the energy--momentum tensor in a linear medium with magnetic and dielectric properties}
\author{Michael E. Crenshaw}
\affiliation{US Army Aviation and Missile Research, Development, and Engineering Center, Redstone Arsenal, AL 35898, USA}
\date{\today}
\begin{abstract}
In a continuum setting, the energy--momentum tensor embodies the
relations between conservation of energy, conservation of linear
momentum, and conservation of angular momentum.
The well-defined total energy and the well-defined total momentum in a
thermodynamically closed system with complete equations of motion are
used to construct the total energy--momentum tensor for a stationary
simple linear material with both magnetic and dielectric properties
illuminated by a quasimonochromatic pulse of light through a
gradient-index antireflection coating.
The perplexing issues surrounding the Abraham and Minkowski momentums
are bypassed by working entirely with conservation principles, the
total energy, and the total momentum.
We derive electromagnetic continuity equations and equations of motion
for the macroscopic fields based on the material four-divergence of the
traceless, symmetric total energy--momentum tensor.
We identify contradictions between the macroscopic Maxwell equations and
the continuum form of the conservation principles.
We resolve the contradictions, which are the actual fundamental issues
underlying the Abraham--Minkowski controversy, by constructing a unified
version of continuum electrodynamics that is based on establishing
consistency between the three-dimensional Maxwell equations for
macroscopic fields, the electromagnetic continuity equations, the
four-divergence of the total energy--momentum tensor, and a four-dimensional
tensor formulation of electrodynamics for macroscopic fields in a
simple linear medium.
\vskip 0.2 cm
\centerline{corresponding author: Michael Crenshaw, email: michael.e.crenshaw4.civ@mail.mil}
\centerline{phone: 1-256-876-3526, fax: 1-256-842-2507}
\end{abstract}
\maketitle
\par
\section{Introduction}
\par
The resolution of the Abraham--Minkowski momentum controversy for the
electromagnetic momentum and the energy--momentum tensor in a linear
medium is multifaceted, complex, and nuanced.
It is not sufficient to simply derive an electromagnetic momentum or
energy--momentum tensor from the macroscopic Maxwell equations, that
much is apparent from any examination of the scientific literature as
the shortcomings of the Abraham momentum \cite{BIAbr} and the 
Minkowski momentum \cite{BIMin} are considerable.
For decades, however, there has been a general consensus in the
scientific literature that the Abraham--Minkowski controversy is
resolved \cite{BIPfei,BIMilBoy,BIKemp,BIBaxL,BIBarL,BIPenHau,BIMikura,
BINelson,BIBPRL,BIObuk,BIGord}:
The Abraham and Minkowski momentum formulas apply to open systems and
each momentum can be supplemented by another component of momentum to
form the total momentum \cite{BIBrev,BIKran}.
Because only the total momentum of a thermodynamically closed
field/matter system is physically meaningful in this paradigm, the total
momentum can be separated into field components, such as the Abraham
and Minkoswski momentums, and matter components in an arbitrary manner.
Although this sophism is ultimately incorrect \cite{BImicro}, it
emphasizes the importance of working with the conserved forms of total
energy and total momentum in order to avoid the ambiguous definitions of
field and material momentums in unclosed physical systems with
incomplete equations of motion \cite{BIKran,BImicro,BICB,BICB2,BISPIE}.
\par
In this article, we {\it carefully} define a thermodynamically closed 
continuum electrodynamic system with complete equations of motion
containing a stationary simple linear medium illuminated by a
plane quasimonochromatic electromagnetic pulse through a gradient-index
antireflection coating and identify the conserved
total energy and the conserved total momentum, thereby extending our
previous work \cite{BICB,BICB2,BISPIE} on dielectrics to a medium with
both magnetic and dielectric properties.
Assuming the validity of the macroscopic Maxwell--Heaviside equations,
the total momentum in a simple linear medium is found using the
law of conservation of linear momentum.
We then populate the total energy--momentum tensor with the densities
of the conserved energy and momentum using the uniqueness property
that certain elements of the tensor correspond to elements of the
four-momentum.
We prove that the uniqueness property directly contradicts the property
that the four-divergence of the energy--momentum tensor generates
continuity equations \cite{BISPIE}.
We retain the uniqueness property and recast the four-divergence of the
energy--momentum tensor in terms of a material four-divergence operator
whose time-like coordinate depends on the refractive
index \cite{BICB,BICB2,BISPIE,BIFinn}.
Once the elements and properties of the energy--momentum tensor are
defined, we derive the continuity equation for the total energy.
The total energy continuity equation is a mixed second-order 
differential equation that we write as first-order equations of
motion for the macroscopic fields.
These equations are mathematically equivalent to the macroscopic
Maxwell--Heaviside equations in that they can be transformed into one
another using vector identities \cite{BIKins,BIFrias}.
It is not that simple.
These results are sufficient for a simple dielectric medium with
$\mu=1$ and the equations of motion for the macroscopic fields
appear in Refs. \cite{BICB,BICB2,BISPIE}.
However, in the more general linear medium considered here, the momentum
continuity equation that is derived from the equations of motion for the
macroscopic fields is not consistent with an energy--momentum tensor
formalism.
In order to resolve this second contradiction, we consider a
quasimonochromatic light pulse traversing a medium consisting of
{\it noninteracting} particles and show that the linear index of
refraction $n$ is $n=1+\xi_e+\xi_m$, where $\xi_e=\chi_e/2$ is the
renormalized electric susceptibility and $\xi_m=\chi_m/2$ is the
renormalized magnetic susceptibility.
Then the set of equations of motion for macroscopic fields that is
consistent with the four-dimensional formulation of the energy
and momentum conservation principles for electrodynamics in the 
continuum limit of a linear medium is: 
\begin{equation}
\frac{n}{c}\frac{\partial {\bf B}}{\partial t} =
\nabla\times {\bf \Pi} -\frac{\nabla n}{n}\times {\bf \Pi}
\label{EQr1.01}
\end{equation}
\begin{equation}
\frac{n}{c}\frac{\partial {\bf \Pi}}{\partial t}
= -\nabla\times {\bf B} 
\label{EQr1.02}
\end{equation}
\begin{equation}
\nabla\cdot {\bf B} =0
\label{EQr1.03}
\end{equation}
\begin{equation}
\nabla\cdot {\bf \Pi} = -\frac{\nabla n}{n}\cdot{\bf \Pi}
\label{EQr1.04}
\end{equation}
where ${\bf B}= \nabla\times{\bf A}$ and
${\bf \Pi}=(n/c)\partial {\bf A}/\partial t$.
While a permittivity $\varepsilon$ and a permeability $\mu$ can still
be defined, the nonlinear disjunction of the linear refractive index
$n=\sqrt{\varepsilon\mu}$ is no longer relevant.
\par
In a continuum setting, the total energy--momentum tensor defines
the relations between the three most important principles of physics:
conservation of energy, conservation of linear momentum, and
conservation of angular momentum.
The inconsistency between the macroscopic Maxwell equations and the
four-dimensional energy--momentum formalism of continuum electrodynamics
is the true and fundamental issue of the Abraham--Minkowski momentum
dilemma.
It would be easy to dismiss our results as contrary to the macroscopic
Maxwell equations, equations that have the status of physical laws.
However to do so would be to ignore contradictions between the 
macroscopic Maxwell equations and the continuum form of the
conservation laws.
This is not the first time that established physical laws
had to be modified when they conflicted with other fundamental
physical principles: Maxwell's modification of the Amp\`ere Law is a
particularly apt example from scientific history.
\par
\section{The Total Energy and the Total Momentum}
\par
A physical theory is an abstract mathematical description of some
portion of a real-world physical process.
The theoretical description is useful to the extent that there
are correspondences between a subset of the theory and a subset of
the behavior of the real-world system \cite{BIRind}.
Real materials are really complicated and we must {\it carefully}
define the material properties and boundary conditions in order to
insure that the mathematical description contains the most important
characteristics while excluding the less significant details.
We define a simple linear material as an isotropic and homogeneous
medium that has a linear magnetic response and a linear dielectric
response to an electromagnetic field that is tuned sufficiently far
from any material resonances that absorption and dispersion are
negligible.
Electrostrictive effects and magnetostrictive effects are also taken
as being negligible \cite{BIMikura}.
Lest there be any doubt about the use of the simple dielectric model
that we described, we note that simplified physical models, like 
elastic collisions, inertial reference frames, weightless and rigid
rods, simple harmonic oscillators, frictionless bearings, plane waves,
the vacuum, and many others, have no existence in the real world
but are nonetheless essential to the progress of theoretical physics.
We are not dealing with any real material, but the idealization
of a simple linear material is necessary in order to develop the
foundational concepts in a clear, concise, convincing, and
incontrovertible manner.
In particular, temporal dispersion is inconsequential for the
arbitrarily long quasimonochromatic electromagnetic field that
is considered here.
\par
A rectangular prism of the simple linear material in free space is
illuminated by a quasimonochromatic pulse of radiation at normal
incidence in the plane-wave limit.
This system is thermodynamically closed with a conserved total energy
and a conserved total momentum but we also require a complete
set of equations of motion in order to obtain well-defined quantities
for the total energy and total momentum.
There are various formulations of the electrodynamics of uniformly
moving media, but a complete treatment of continuum electrodynamics in
moving media is far from settled \cite{BIPenHau}.
We also have to recognize that the material is accelerating due to a
radiation surface pressure that can be attributed to the partial
reflection of the incident radiation.
In this work, the prism of material is covered with a thin
gradient-index antireflection coating, not an interference
antireflection coating, that makes reflections and the acceleration from
radiation surface pressure negligible allowing the material to be
regarded as stationary in the laboratory frame of reference.
The assumption of a rare vapor, by Gordon \cite{BIGord} for example,
accomplishes the same purpose of making reflections negligible, but
then requires justification to apply the results to a material with
a non-perturbative index of refraction.
There is nothing unusual, mysterious, nefarious, or confusing about the
conditions described above because the use of a gradient-index
antireflection coating on a homogeneous linear material is required,
and has {\it always} been required, for a rigorous application of
the Maxwell equations to solids.
Still, Maxwell's equations are almost always applied to
moving and accelerating materials without adequate justification.
That is not to say that the use of the macroscopic Maxwell equations
cannot be justified in most cases, but it is especially important to
justify the use of the macroscopic Maxwell equations when invoking
conservation properties.
\par
Classical continuum electrodynamics is founded on the macroscopic
Maxwell equations, so that is where we start our work.
The common macroscopic Maxwell--Heaviside equations
\begin{equation}
\nabla\times {\bf E}
=- \frac{\mu}{c}\frac{\partial {\bf H}}{\partial t}
\label{EQr2.01}
\end{equation}
\begin{equation}
\nabla\times {\bf H}
= \frac{\varepsilon}{c}\frac{\partial {\bf E}}{\partial t}
\label{EQr2.02}
\end{equation}
\begin{equation}
\nabla\cdot{\bf B} = 0
\label{EQr2.03}
\end{equation}
\begin{equation}
\nabla\cdot \varepsilon{\bf E} = 0
\label{EQr2.04}
\end{equation}
are the complete equations of motion for the macroscopic fields in a
stationary simple linear medium.
Here, ${\bf E}$ is the electric field,
${\bf B}$ is the magnetic field,
${\bf H}={\bf B}/\mu$ is the auxiliary magnetic field,
$\varepsilon$ is the electric permittivity,
$\mu$ is the magnetic permeability,
and $c$ is the speed of light in the vacuum.
{\it The medium is explicitly required to be stationary} because the 
macroscopic Maxwell--Heaviside equations,
Eqs.~(\ref{EQr2.01})--(\ref{EQr2.04}), are not complete for moving or
accelerating materials.
The electric and magnetic fields can be defined in terms of the vector 
potential in the usual manner as
\begin{equation}
{\bf E}= -\frac{1}{c}\frac{\partial{\bf A}}{\partial t}
\label{EQr2.05}
\end{equation}
\begin{equation}
{\bf B}= \nabla\times {\bf A} \, ,
\label{EQr2.06}
\end{equation}
working in the Coulomb gauge.
Then, the propagation of an electromagnetic field through free-space and
into a simple linear medium is described by the wave equation,
\begin{equation}
\nabla\times(\nabla\times{\bf A})
+\frac{n^2}{c^2}\frac{\partial^2{\bf A}}{\partial t^2}
=\frac{\nabla\mu}{\mu}\times(\nabla\times{\bf A}) \, ,
\label{EQr2.07}
\end{equation}
assuming the validity of the macroscopic Maxwell--Heaviside equations,
Eqs.~(\ref{EQr2.01})--(\ref{EQr2.04}).
Here, $n({\bf r})=\sqrt{\mu\varepsilon}$ is the spatially slowly varying
linear refractive index.
The spatial variation of the index and permeability is limited to
a narrow transition region in which these quantities change gradually
from the vacuum values to the nominal material properties.
We write the vector potential in terms of a slowly varying
envelope function and a carrier wave as
\begin{equation}
{\bf A}({\bf r},t)= \frac{1}{2} \left (
{\bf \tilde A}({\bf r},t)e^{-i(\omega_d t-{\bf k}_d\cdot{\bf r})}
+ {\bf \tilde A}^* ({\bf r},t)e^{i(\omega_d t-{\bf k}_d\cdot{\bf r})}
\right ) \, ,
\label{EQr2.08}
\end{equation}
where $\tilde A$ is a slowly varying function of ${\bf r}$ and $t$,
${\bf k}_d=(n\omega_d/c) {\bf \hat e}_{\bf k}$ is the wave vector 
that is associated with the center frequency of the field $\omega_d$,
and ${\bf \hat e}_{\bf k}$ is a unit vector in the direction of propagation.
\par
Figure 1 shows a one-dimensional representation of the slowly varying
amplitude of the plane incident field
$\tilde A_i(z)=({\bf\tilde A}(z,t_0)\cdot{\bf\tilde A}^*(z,t_0))^{1/2}$
about to enter the simple linear medium with index $n=1.386$ and 
permeability $\mu=1.2$ through a gradient-index antireflection
coating.
\begin{figure}
\includegraphics[scale=0.30]{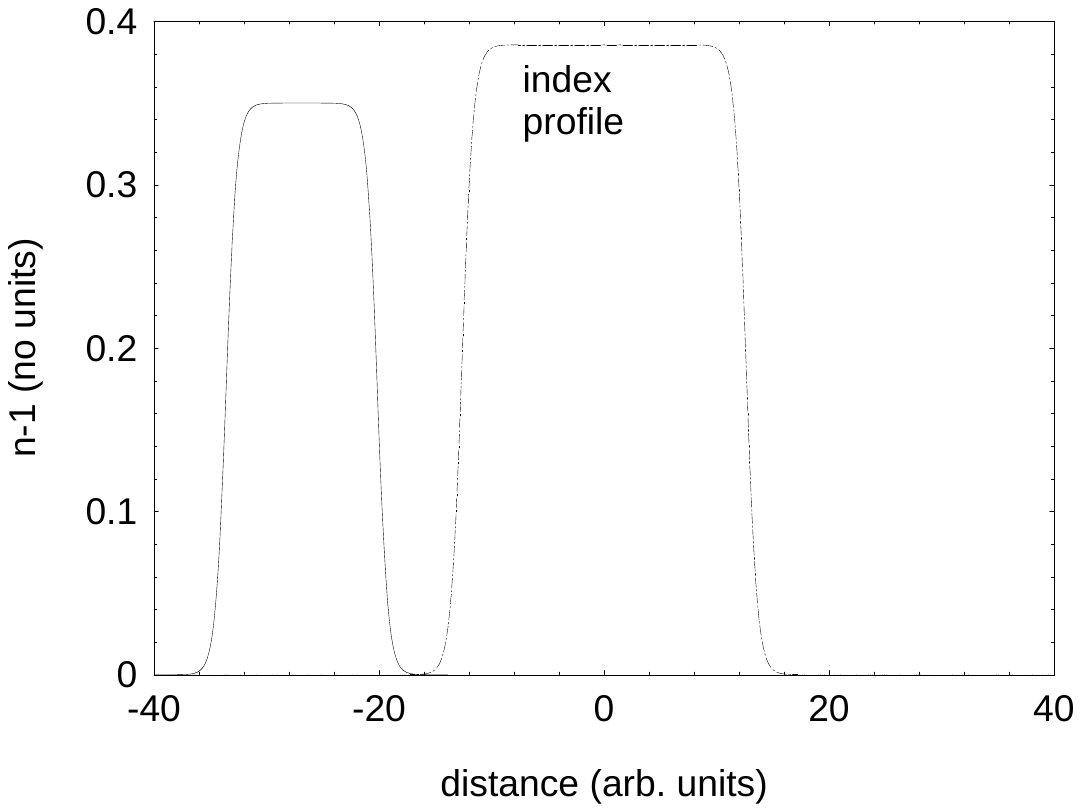}
\caption{Incident field (amplitude arbitrary) just before it enters
the linear medium.}
\label{fig1}
\end{figure}
The gradient that has been applied to the index has also been applied
to the permeability as a matter of convenience.
Figure 2 presents a time-domain numerical solution of the wave equation
at a later time $t_1$ when the refracted field
$\tilde A_t(z)=({\bf\tilde A}(z,t_1)\cdot{\bf \tilde A}^*(z,t_1))^{1/2}$
is entirely inside the medium.
\begin{figure}
\includegraphics[scale=0.30]{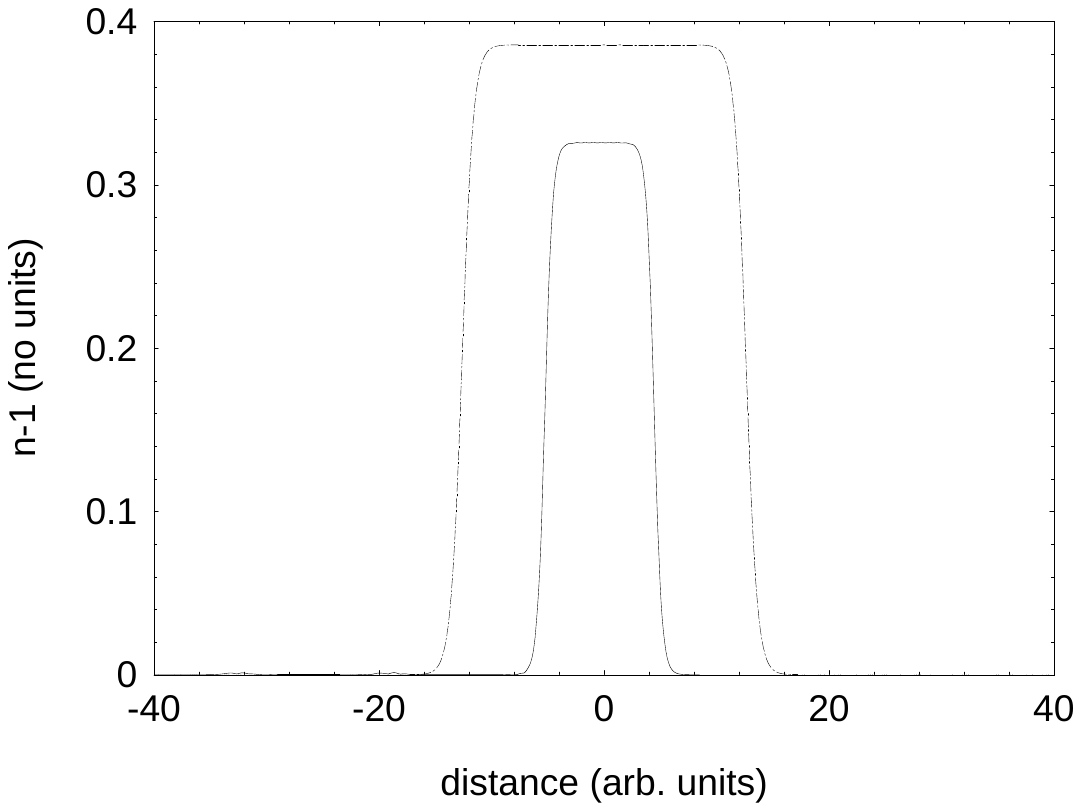}
\caption{Refracted field entirely within the linear medium.}
\label{fig2}
\end{figure}
The pulse has not propagated as far as it would have propagated in the
vacuum due to the reduced speed of light $c/n$ in the material.
In addition, the spatial extent of the refracted pulse in the medium
is $w_t=w_i/n$ in terms of the width $w_i$ of the incidence pulse due
to the reduced speed of light.
As shown in Fig. 2, the amplitude of the refracted field is
${\tilde A}_t=\sqrt{{\mu}/{n}}{\tilde A}_i $, that is, the amplitude of
the incident field scaled by $\sqrt{\mu/n}$.
Using these relations we can construct the temporally invariant quantity
\begin{equation}
U_{total}=\int_{\sigma}
\frac{n}{2}
\left (\sqrt{\frac{n}{\mu}}\frac{\omega_d}{c}\tilde A \right )
\left (\sqrt{\frac{n}{\mu}}\frac{\omega_d}{c}\tilde A \right )^*
dv
\label{EQr2.09}
\end{equation}
and note that $U_{total}$ is conserved when the integration is
performed over a region $V$ that contains all of the electromagnetic
field that is present in the system.
In that case, the region of integration $V$ is extended to all
space, $\sigma$.
The pulse that is used in Fig. 1 has a generally rectangular shape in
order to facilitate a graphical interpretation of pulse width and 
integration under the field envelope.
It can be shown by additional numerical solutions of the wave equation
that the relations described above are quite general in terms of the
permeability and the refractive index, as well as the shape and
amplitude of a quasimonochromatic field.
\par
For a stationary linear medium, the total energy is the electromagnetic
energy.
Although a material may possess many forms of energy, we intend total
energy to mean all of the energy that impacts the dynamics or
electrodynamics of the model system with the specified conditions.
In particular, kinetic energy is excepted by the requirement for the
material to remain stationary.
We assume that the electromagnetic energy density for a stationary
simple linear medium is
\begin{equation}
\rho_{e}=\frac{1}{2}\frac{1}{\mu}
\left (n^2{\bf E}^2+{\bf B}^2\right )dv \, ,
\label{EQr2.10}
\end{equation}
as inferred from the Poynting theorem.
Again extending the region of integration $V$ that contains all fields
present to integration over all space $\sigma$, the electromagnetic
energy,
\begin{equation}
U_{e}=\int_{\sigma} \rho_e dv
=\int_{\sigma} \frac{1}{2}\frac{1}{\mu}
\left ( n^2{\bf E}^2+{\bf B}^2\right ) dv \, ,
\label{EQr2.11}
\end{equation}
is the total energy of the closed system containing a stationary simple
linear medium illuminated by a quasimonochromatic pulse of light.
Using Eqs.~(\ref{EQr2.05}) and (\ref{EQr2.06}) to eliminate the electric
and magnetic fields in Eq.~(\ref{EQr2.11}), we obtain the energy formula
\begin{equation}
U_{e}=\int_{\sigma} \frac{1}{2}\frac{1}{\mu}
\left (
\left ( \frac{n}{c}\frac{\partial{\bf A}}{\partial t}\right )^2
+(\nabla\times{\bf A})^2
\right ) dv \, .
\label{EQr2.12}
\end{equation}
Employing the expression for the vector potential in terms of slowly
varying envelope functions and carrier waves, Eq.~(\ref{EQr2.08}),
results in
\begin{equation}
U_{e}=\int_{\sigma}\frac{\omega_d^2n^2}{4\mu c^2}
\left (
\tilde A\tilde A^*
-\tilde A^{2} e^{-2i(\omega_d-k_dz) t}
+c.c.
\right ) dv \, .
\label{EQr2.13}
\end{equation}
Applying a time average allows the double frequency
terms to be neglected and shows the equality of Eqs.~(\ref{EQr2.09})
and (\ref{EQr2.12}),
\begin{equation}
U_{e}=\int_{\sigma}\frac{\omega_d^2n^2}{2\mu c^2} |\tilde A|^2 dv
=U_{total} \, .
\label{EQr2.14}
\end{equation}
We have theoretical confirmation that our numerical $U_{total}$,
Eq.~(\ref{EQr2.09}), is the total energy of the system and that it is
conserved.
More importantly, we have an interpretation of the continuum
electrodynamic energy formula, Eq.~(\ref{EQr2.11}), in term of what
happens to the shape and amplitude of the electromagnetic field as it
propagates into the medium.
\par
In the first half of the last century, there was an ongoing theoretical
and experimental effort devoted to determining the correct description
of the momentum of light propagating through a linear medium 
\cite{BIPfei,BIMilBoy,BIKemp,BIBaxL,BIBarL}.
At various times, the issue was resolved in favor of the Abraham
momentum
\begin{equation}
{\bf G}_A=\int_{\sigma}\frac{{\bf E}\times{\bf H}}{c} dv
\label{EQr2.15}
\end{equation}
and at other times the momentum was found to be the Minkowski momentum 
\begin{equation}
{\bf G}_M=\int_{\sigma}\frac{{\bf D}\times{\bf B}}{c} dv \, .
\label{EQr2.16}
\end{equation}
Since Penfield and Haus \cite{BIPenHau} showed, in 1967, that neither
the Abraham momentum nor the Minkowski momentum is the total momentum,
the resolution of the Abraham--Minkowski controversy has been that the
momentum for an arbitrary or unspecified field subsystem must be
supplemented by an appropriate material momentum to obtain the total
momentum \cite{BIPfei,BIMilBoy,BIKemp,BIBaxL,BIBarL,
BIPenHau,BIMikura,BINelson,BIBPRL,BIObuk,BIGord,BIBrev,BIKran}.
Then, it must be demonstrated that the total momentum, so constructed,
is actually conserved in a thermodynamically closed system, as that is
usually not the case.
Although continuity equations are often called conservation laws, the
derivation of a result that has the outward appearance of a continuity
equation is not sufficient to prove conservation.
\par
Given the arbitrariness of the field momentum and the model dependence
of the material momentum that go into a construction of the total
momentum, the only convincing way to determine the total momentum is
by the application of conservation principles.
Then conservation of the total energy is all that is needed
to prove that the total momentum \cite{BIGord,BICB,BICB2,BISPIE}
\begin{equation}
{\bf G}_{total}
=\int_{\sigma} \frac{1}{\mu} \frac{ n{\bf E}\times{\bf B}}{c} dv
=\int_{\sigma} \frac{ n{\bf E}\times{\bf H}}{c} dv
\label{EQr2.17}
\end{equation}
is conserved in our closed system.
In the limit of slowly varying plane waves, we have the conserved vector
quantity
\begin{equation}
{\bf G}_{total}= \int_{\sigma}\frac{\omega_d^2n^2}{2\mu c^3}
|\tilde A|^2 dv {\bf e}_{\bf k} 
= \frac{U_{total}}{c} {\bf e}_{\bf k}
\label{EQr2.18}
\end{equation}
that is equal to the conserved total energy, Eq.~(\ref{EQr2.14}),
divided by a constant factor of $c$ and multiplied by a unit vector
${\bf e}_{\bf k}$.
In addition to being conserved, the total momentum, Eq.~(\ref{EQr2.17}),
is unique.
There can be no other non-trivial conserved macroscopic momentum in
terms of the vector product of electric and magnetic fields in our model
system because these would contain different combinations of the 
material properties.
The momentum formula, Eq.~(\ref{EQr2.17}), was originally derived for 
a dielectric in 1973 by Gordon \cite{BIGord} who combined the Abraham
momentum for the field with a material momentum that was obtained from
integrating the Lorentz force on the microscopic constituents of matter.
Although, Gordon's derivation is incorrect and incomplete \cite{BImicro},
conservation of the total momentum is definitive and the Gordon
momentum is uniquely conserved in a thermodynamically closed continuum
electrodynamic system containing a stationary simple linear medium.
The Gordon momentum is actually composed of the momentum of the electric,
magnetic, and polarization fields \cite{BImicro}.
We can associate the polarization field with the material, but apart
from a semantic artifice, the momentum, like the energy, is entirely 
electromagnetic in nature.
\par
\section{Total Energy--Momentum Tensor}
\par
A tensor is a very special mathematical object and a total
energy--momentum tensor is subject to even more stringent conditions.
{\it Any} four continuity equations, or equations that have the outward
appearance of being continuity equations, can be combined using linear
algebra to form a matrix differential equation.
Hence, the proliferation of matrices that have been purported to be the
energy--momentum tensor for a system containing an electromagnetic pulse
and a linear medium.
In this section, we construct the unique total energy--momentum tensor
and the tensor continuity equation for a stationary simple linear medium
using the conservation, symmetry, trace, and divergence conditions that
must be satisfied.
\par
For an unimpeded (force-free) flow, conservation of the components of
the total four-momentum $(U,{\bf G})$ \cite{BILL}
\begin{equation}
U=\int_{\sigma} T^{00} dv
\label{EQr3.01}
\end{equation}
\begin{equation}
G^{i}=\frac{1}{c}\int_{\sigma} T^{i0} dv
\label{EQr3.02}
\end{equation}
uniquely determines the first column of the total energy--momentum
tensor.
We use the convention that Roman indices from the middle of the
alphabet, like $i$, run from 1 to 3 and Greek indices belong
to $\{0,1,2,3\}$.
As in Section II, the region of integration has been extended to
all-space, $\sigma$.
Conservation of angular momentum in a closed system imposes the diagonal
symmetry condition \cite{BILL}
\begin{equation}
T^{\alpha\beta}=T^{\beta\alpha}
\label{EQr3.03}
\end{equation}
and uniquely determines the first row of the total energy--momentum
tensor based on the uniqueness of the first column.
Applying the conservation of the total energy and total momentum, we can
construct the total energy--momentum tensor
$$
T^{\alpha\beta}=
$$
\begin{equation}
\left [
\begin{matrix}
(n^2{\bf E}^2+{\bf B}^2)/(2\mu)
&(n{\bf E}\times{\bf H})_1
&(n{\bf E}\times{\bf H})_2
&(n{\bf E}\times{\bf H})_3
\cr
(n{\bf E}\times{\bf H})_1
&W_{11}
&W_{12}
&W_{13}
\cr
(n{\bf E}\times{\bf H})_2
&W_{21}
&W_{22}
&W_{23}
\cr
(n{\bf E}\times{\bf H})_3
&W_{31}
&W_{32}
&W_{33}
\cr
\end{matrix}
\right ] 
\label{EQr3.04}
\end{equation}
from elements that are obtained by equating
Eq.~(\ref{EQr2.11}) to Eq.~(\ref{EQr3.01})
and equating Eq.~(\ref{EQr2.17}) to Eq.~(\ref{EQr3.02}).
The elements of the Maxwell stress-tensor $W$ are yet to be
specified.
\par
Our approach here is very different from the usual technique of
constructing the energy--momentum tensor from the electromagnetic
continuity equations.
The energy continuity equation is known to be given by
Poynting's theorem
\begin{equation}
\frac{1}{c}\frac{\partial \rho_e}{\partial t}+
\nabla\cdot \left ( {\bf E}\times{\bf H}\right ) =0 \, .
\label{EQr3.05}
\end{equation}
However if we use Poynting's theorem and the
tensor energy continuity law
\begin{equation}
\partial_{\beta}T^{0\beta}=0
\label{EQr3.06}
\end{equation}
to populate the first row of the energy--momentum tensor, then
Eq.~(\ref{EQr3.02})
becomes
\begin{equation}
G^i=\frac{1}{c}\int_{\sigma} ({\bf E}\times{\bf H})_i dv
\label{EQr3.07}
\end{equation}
by symmetry, Eq.~(\ref{EQr3.03}).
This result is contraindicated because
\begin{equation}
{\bf G}
=\frac{1}{c}\int_{\sigma}\frac{\omega_d^2n}{2\mu c^2}|\tilde A|^2
{\bf e}_{\bf k}dv
=\frac{U_{total}}{nc} {\bf e}_{\bf k}
\label{EQr3.08}
\end{equation}
is not temporally invariant as the pulse travels from the vacuum
into the medium.
The practice has been to ignore the condition that the components of the
four-momentum, Eqs.~(\ref{EQr3.01}) and (\ref{EQr3.02}), be conserved
and use the components of the Poynting vector, along with the
electromagnetic energy density, to populate the first row of the
energy--momentum tensor using the tensor energy continuity law,
Eq.~(\ref{EQr3.06}).
\par
We regard the conservation properties of the total energy--momentum
tensor as fundamental and these conservation properties conclusively
establish Eq.~(\ref{EQr3.04}) as the form of the energy--momentum
tensor.
Applying the tensor continuity equation, Eq.~(\ref{EQr3.06}), to the
energy--momentum tensor, Eq.~(\ref{EQr3.04}), we obtain an energy
continuity equation
\begin{equation}
\frac{1}{c}\frac{\partial \rho_e}{\partial t}+
\nabla\cdot \left ( n {\bf E}\times{\bf H}\right ) =0 
\label{EQr3.09}
\end{equation}
that is inconsistent with the Poynting theorem, Eq.~(\ref{EQr3.05}).
In the limit of slowly varying plane waves, we obtain an obvious
contradiction
\begin{equation}
\frac{n^2\omega_d^2}{2\mu c^3}
\left (
{\bf\tilde A}\frac{\partial{\bf\tilde A}^*}{\partial t}+c.c.
\right )
- \frac{n^3\omega_d^2}{2\mu c^3} 
\left (
{\bf\tilde A}\frac{\partial{\bf\tilde A}^*}{\partial t}+c.c.
\right )
= 0
\label{EQr3.10}
\end{equation}
from the energy continuity equation, Eq.~(\ref{EQr3.09}).
\par
We have identified the essential contradiction of the Abraham--Minkowski
controversy:
Absent the uniqueness that accompanies conservation of total energy and
total momentum, the components of the first row and column of the
energy--momentum tensor are essentially arbitrary thereby rendering the
energy--momentum tensor meaningless.
Yet, the energy continuity equation that is obtained from the
four-divergence of an energy--momentum tensor that is populated with 
the densities of the conserved energy and momentum quantities is
demonstrably false.
We are at an impasse with the existing theory that requires two
contradictory conditions to be satisfied \cite{BISPIE}.
\par
The resolution of this contradiction cannot be found within the 
formal system of continuum electrodynamics as it currently exists.
Therefore we make an $ansatz$, based on the reduced speed of light in
the medium, $c/n$, that the continuity equations are generated by 
\begin{equation}
\bar\partial_{\beta}T^{\alpha\beta}= f_{\alpha} \, ,
\label{EQr3.11}
\end{equation}
where $\bar\partial_{\beta}$ is the material four-divergence
operator \cite{BICB,BICB2,BISPIE,BIFinn}
\begin{equation}
\bar\partial_{\beta}=\left ( \frac{n}{c}\frac{\partial}{\partial t},
\frac{\partial}{\partial x},\frac{\partial}{\partial y},
\frac{\partial}{\partial z} \right ) \, .
\label{EQr3.12}
\end{equation}
The continuity equation, Eq.~(\ref{EQr3.11}), is generalized in a
form that allows a source/sink $f_{\alpha}$ of energy and the
components of momentum \cite{BICB2}.
Reference \cite{BIxxx} provides a solid theoretical justification
for the {\it ansatz} of Eqs.~(\ref{EQr3.11}) and (\ref{EQr3.12}).
Applying the tensor continuity law, Eq.~(\ref{EQr3.11}), to the
energy--momentum tensor, Eq.~(\ref{EQr3.04}), we obtain the energy
continuity equation
\begin{equation}
\frac{n}{c}\frac{\partial \rho_e}{\partial t}+
\nabla\cdot \left ( n{\bf E}\times{\bf H}\right ) =f_0 \, .
\label{EQr3.13}
\end{equation}
Now, let us re-consider the Poynting theorem.
We multiply Poynting's theorem, Eq.~(\ref{EQr3.05}), by $n$ and use a
vector identity to commute the index of refraction with the divergence
operator to obtain \cite{BICB}
\begin{equation}
\frac{n}{c}\frac{\partial \rho_e}{\partial t}+
\nabla\cdot \left ( n{\bf E}\times{\bf H}\right )
= \frac{\nabla n}{n} \cdot ( n{\bf E}\times{\bf H} ) \, .
\label{EQr3.14}
\end{equation}
The two energy continuity equations, Eqs.~(\ref{EQr3.13}) and
(\ref{EQr3.14}), are equal if we identify the source/sink of energy as 
\begin{equation}
f_0=\frac{\nabla n}{n} \cdot ( n{\bf E}\times{\bf H}) \, .
\label{EQr3.15}
\end{equation}
Then an inhomogeneity in the index of refraction, $\nabla n\ne 0$, is
associated with work done by the field on the material or work done
on the field by the interaction with the material \cite{BICB2}.
Obviously, the gradient of the index of refraction in our
anti-reflection coating must be sufficiently small that the 
work done is perturbative.
\par
The Poynting's theorem that is found in Eq.~(\ref{EQr3.14})
is a mixed second-order vector differential equation.
It can be separated into mixed first-order vector differential
equations for the macroscopic electric and magnetic fields
\begin{equation}
\frac{n}{c}\frac{\partial {\bf B}}{\partial t}
=
-\nabla\times n{\bf E} +\frac{\nabla n}{n}\times n{\bf E}
\label{EQr3.16}
\end{equation}
\begin{equation}
\frac{n}{c}\frac{\partial n{\bf E}}{\partial t}
= \mu \nabla\times {\bf H}.
\label{EQr3.17}
\end{equation}
The remaining task is to show that the material four-divergence of
the energy--momentum tensor is a faithful representation of the 
electromagnetic continuity equations.
We multiply Eq.~(\ref{EQr3.16}) by ${\bf H}$ and multiply
Eq.~(\ref{EQr3.17}) by $n{\bf E}$.
The resulting equations are summed to produce an energy continuity
equation
\begin{equation}
\frac{n}{c}\frac{\partial \rho_e}{\partial t}+
\nabla\cdot \left ( n{\bf E}\times{\bf H}\right )
= \frac{\nabla n}{n} \cdot ( n{\bf E}\times{\bf H} ) 
\label{EQr3.18}
\end{equation}
that is precisely the same as Eq.~(\ref{EQr3.14}).
To obtain the total momentum continuity equation, we substitute
Eqs.~(\ref{EQr3.16}) and (\ref{EQr3.17}) into the material timelike
derivative of the total momentum density
\begin{equation}
\frac{n}{c}\frac{\partial {\bf g}_{total}}{\partial t}
=
\frac{1}{\mu}\frac{n}{c^2}
\frac{\partial n{\bf E}}{\partial t}\times{\bf B}+
n{\bf E}\times\frac{1}{\mu}
\frac{n}{c^2}\frac{\partial{\bf B}}{\partial t} \, 
\label{EQr3.19}
\end{equation}
where the momentum density ${\bf g}_{total}$ is the integrand of
Eq.~(\ref{EQr2.17}).
The momentum continuity equation is
$$
\frac{n}{c}\frac{\partial {\bf g}_{total}}{\partial t} =
\frac{1}{c\mu}\Bigg [
(\nabla\times n{\bf E})\times n{\bf E}
$$
\begin{equation}
+\mu^2 (\nabla\times {\bf H})\times {\bf H}
+\frac{\nabla n}{n}(n{\bf E})^2
-\left (n{\bf E}\cdot\frac{\nabla n}{n}\right ) n{\bf E}
\Bigg ].
\label{EQr3.20}
\end{equation}
For a dielectric medium, $\mu=1$, we are done and this case was treated
in Refs. \cite{BICB,BICB2,BISPIE}.
However, if $\mu\ne 1$ then the momentum continuity equation,
Eq.~(\ref{EQr3.20}), is not expressible as part of the tensor continuity
equation, Eq.~(\ref{EQr3.11}), without additional transformations.
\par
There is a scientific record of algebraic transformations of the energy
and momentum continuity equations.
Abraham \cite{BIAbr} was apparently the first to pursue this approach,
deriving a variant form of the continuity equation for the Minkowski
momentum in which the time-derivative of the Minkowski momentum is
split into the temporal derivative of the Abraham momentum and a
fictitious Abraham force.
Kinsler, Favaro, and McCall \cite{BIKins} discuss various
transformations of Poynting's theorem and the resulting differences
in the way various physical processes are expressed.
Frias and Smolyakov \cite{BIFrias} did the same for transformations
of the momentum continuity equation.
It appears to have gone unrecognized, however, that the different
expression of physical processes in the electromagnetic continuity
equations has to carry over into the Maxwell equations of
motion for the macroscopic fields, Eqs.~(\ref{EQr3.16}) and
(\ref{EQr3.17}).
Specifically, the energy and momentum continuity equations
are components of the tensor continuity equation, Eq.~(\ref{EQr3.11}),
and cannot be separately transformed.
We could, for example, commute the magnetic permeability with the curl
operator in the Maxwell--Amp\`ere Law, Eq.~(\ref{EQr3.17}), in order to
allow the momentum continuity law, Eq.~(\ref{EQr3.20}) to be
expressed in the form of a continuity equation.
However, the resulting energy continuity law would run afoul of the
uniqueness condition on the elements of the first column of the
energy--momentum tensor.
\par
The total energy--momentum tensor, Eq.~(\ref{EQr3.04}), was constructed
using conservation principles and it is inarguably correct as long as
the formula for the energy, Eq.~(\ref{EQr2.11}), is correct.
For a linear medium with a full or partial magnetic response, however,
we have demonstrated that the continuity equations,
Eqs.~(\ref{EQr3.18}) and (\ref{EQr3.20}), that were constructed from
the field equations, Eq.~(\ref{EQr3.16}) and (\ref{EQr3.17}), are not
consistent with the tensor continuity equation, Eq.~(\ref{EQr3.11}).
Again, we are confronted with a situation of contradiction that we
cannot derive our way out of and we must justify a correction based
on an analysis of the physical model.
\par
\section{The Linear Index of Refraction}
\par
Consider a rectangular volume of space $V=Ad$, where $A$ is the
cross-sectional area perpendicular to the direction of an incident 
electromagnetic pulse of radiation in the plane-wave limit.
The volume is separated into halves by a thin transparent barrier at
$d/2$.
Half of the volume is filled with a vapor with refractive index $n_1$
and the other half of the volume is filled with a vapor with refractive
index $n_2$.
We define a material variable $\xi_i=n_i-1$.
Then the time that it takes a light pulse to traverse the volume is
$(\xi_1+\xi_2)d/(2c)$ greater than the time it takes light to travel
the same distance in a vacuum.
Note that reflections do not affect the time-of-flight of the pulse,
only the amount of light that is transmitted.
The time-of-flight of the light pulse does not change when the barrier
is removed.
The atoms of each vapor are noninteracting and there is no physical
process that would affect the time-of-flight of the light pulse as
the two species of atoms diffuse into each other.
Generalizing to an arbitrary number of components, $s$, the linear
refractive index of a mixture of vapors obeys a superposition principle 
\begin{equation}
n=1+\sum_{k=1}^s \xi_k .
\label{EQr4.01}
\end{equation}
As long as the atoms are noninteracting, the superposition principle
holds whether the vapors are composed of atoms that behave as
electric dipoles, magnetic dipoles, or as a mixture of both.
That is, 
\begin{equation}
n=1+\sum_{i=1}^p \xi_i +\sum_{j=1}^q \xi_j 
\label{EQr4.02}
\end{equation}
for $p$ dielectric components and $q$ magnetic components.
Then, $n=1+\xi_e+\xi_m$,
where $\xi_e=\chi_e/2$ is the renormalized electric susceptibility and 
$\xi_m=\chi_m/2$ is the renormalized magnetic susceptibility.
It should be noted that this is not the first time that the refractive
index was derived as a linear superposition of the electric and magnetic
dipole densities \cite{BIWardWebb}.
It is also common in the literature to invoke the low-density limit so
that $n=1+\xi_e+\xi_m$ can be used in place of
$n=\sqrt{(1+\chi_e)(1+\chi_m)}$, Ref.~\cite{BIGord}, for example 
\cite{BImicro}.
Near-dipole--dipole interactions \cite{BIChuck}, and other interactions
between atoms, require a more detailed and model-dependent treatment
that is outside the scope of the current work.
\par
Now that the linear refractive index is truly linear, we can regard the
permittivity and permeability as anachronisms.
The refractive index carries the material effects and we can eliminate
the permeability and permittivity from the equations of motion for the
macroscopic fields, Eqs.~(\ref{EQr3.16}) and (\ref{EQr3.17}), such that
\begin{equation}
\frac{n}{c}\frac{\partial {\bf B}}{\partial t}
=
\nabla\times {\bf \Pi} -\frac{\nabla n}{n}\times {\bf \Pi}
\label{EQr4.03}
\end{equation}
\begin{equation}
\frac{n}{c}\frac{\partial {\bf \Pi}}{\partial t}
= -\nabla\times {\bf B} \, ,
\label{EQr4.04}
\end{equation}
where ${\bf \Pi}=(n/c)\partial{\bf A}/\partial t$.
The equations of motion for the macroscopic fields,
Eqs.~(\ref{EQr4.03}) and (\ref{EQr4.04}),
can be combined in the usual manner to write an energy
continuity equation
\begin{equation}
\frac{n}{c}\frac{\partial }{\partial t} \left [
\frac{1}{2}\left ( {\bf \Pi}^2+{\bf B}^2\right ) 
 \right ]+
\nabla\cdot \left ( {\bf B}\times{\bf \Pi}\right )
=\frac{\nabla n}{n} \cdot({\bf B}\times{\bf \Pi}) 
\label{EQr4.05}
\end{equation}
in terms of a total energy density
\begin{equation}
\rho_{total}=\frac{1}{2}\left ( {\bf \Pi}^2+{\bf B}^2\right )  \, ,
\label{EQr4.06}
\end{equation}
a total momentum density
\begin{equation}
{\bf g}_{total}=\frac{{\bf B}\times{\bf \Pi}}{c}  \, ,
\label{EQr4.07}
\end{equation}
and a power density
\begin{equation}
p_{total}=\frac{\nabla n}{n} \cdot({\bf B}\times{\bf \Pi}) \, .
\label{EQr4.08}
\end{equation}
Substituting Eqs.~(\ref{EQr4.03}) and (\ref{EQr4.04}) into the 
material timelike derivative of the total momentum,
Eq.~(\ref{EQr4.07}), the momentum continuity equation becomes
\begin{equation}
\frac{n}{c}\frac{\partial {\bf g}_{total}}{\partial t}+
\frac{1}{c}\nabla\cdot{\bf W}=
\frac{1}{c}\frac{\nabla n}{n}{\bf \Pi}^2 \, ,
\label{EQr4.09}
\end{equation}
where the Maxwell stress-tensor $W$ is 
\begin{equation}
W_{ij}=-\Pi_i\Pi_j-B_iB_j+
\frac{1}{2} ({\bf \Pi}^2+{\bf B}^2)\delta_{ij} \, .
\label{EQr4.10}
\end{equation}
We can construct the total energy--momentum tensor
\begin{equation}
T^{\alpha\beta}=
\left [
\begin{matrix}
({\bf \Pi}^2+{\bf B}^2)/2  &({\bf B}\times{\bf \Pi})_1 
&({\bf B}\times{\bf \Pi})_2  &({\bf B}\times{\bf \Pi})_3
\cr
({\bf B}\times{\bf \Pi})_1    &W_{11}      &W_{12}      &W_{13}  
\cr
({\bf B}\times{\bf \Pi})_2    &W_{21}      &W_{22}      &W_{23}     
\cr
({\bf B}\times{\bf \Pi})_3    &W_{31}      &W_{32}      &W_{33}   
\cr
\end{matrix}
\right ]
\label{EQr4.11}
\end{equation}
from the homogeneous part of the new electromagnetic continuity
equations, Eqs.~(\ref{EQr4.05}) and (\ref{EQr4.09}).
The total energy--momentum tensor, Eq.~(\ref{EQr4.11}), is entirely 
electromagnetic in character {\cite{BImicro} and there is no need for
a supplemental dust energy--momentum tensor for the movement of the
material {\cite{BIPfei}.
The Maxwell--Amp\`ere Law, Eq.~(\ref{EQr4.04}), can be written in
terms of the vector potential as a wave equation
\begin{equation}
\nabla\times(\nabla\times{\bf A})
+\frac{n^2}{c^2}\frac{\partial^2{\bf A}}{\partial t^2} =0 \, ,
\label{EQr4.12}
\end{equation}
without the inhomogeneous part found in Eq.~(\ref{EQr2.07}).
We also correct the electromagnetic, or total, energy,
Eq.~(\ref{EQr2.11}),
\begin{equation}
U_{e}=\int_{\sigma} \rho_e dv
=\int_{\sigma} \frac{1}{2}
\left ({\bf \Pi}^2+{\bf B}^2\right ) dv \, ,
\label{EQr4.13}
\end{equation}
and the total momentum, Eq.~(\ref{EQr2.17}),
\begin{equation}
{\bf G}_{total}
=\int_{\sigma} \frac{ {\bf B}\times{\bf \Pi}}{c} dv 
\label{EQr4.14}
\end{equation}
using the total energy density, Eq.~(\ref{EQr4.06}), and
the total momentum density, Eq.~(\ref{EQr4.07}).
\par
The ${\{\bf E}$, ${\bf D}$, ${\bf B}$, ${\bf H}\}$ paradigm of
classical continuum electrodynamics is altogether broken.
That much was obvious from the conservation of the Gordon form of 
momentum in Refs.~\cite{BIGord} and \cite{BICB}, also in
Eq.~(\ref{EQr2.17}).
The Abraham and Minkowski momentums are expressible solely in terms
of the ${\{\bf E}$, ${\bf D}$, ${\bf B}$, ${\bf H}$\} fields, while the
Gordon momentum depends on the magnetic field ${\bf B}$ and the peculiar
macroscopic field ${\bf \Pi}=-n{\bf E}$.
This is likely a contributing factor to the longevity of the
Abraham--Minkowski controversy.
The macroscopic field equations, Eqs.~(\ref{EQr4.03}) and
(\ref{EQr4.04}), depend only on the fields ${\bf \Pi}$ and ${\bf B}$.
Consequently, continuum electrodynamics can be reformulated in terms
of a single pair of fields $\{{\bf \Pi},{\bf B}\}$ and a single
field-strength tensor \cite{BIxxx}
\begin{equation}
F^{\alpha\beta}=
\left [
\begin{matrix}
 0        &\Pi_x      &\Pi_y      &\Pi_z
\cr
-\Pi_x     &0           &-B_z        &B_y     
\cr
-\Pi_y     &B_z         &0           &-B_x       
\cr
-\Pi_z     &-B_y        &B_x         &0        
\cr
\end{matrix}
\right ] \, .
\label{EQb4.15}
\end{equation}
Having eliminated $\varepsilon$ and $\mu$, the fields ${\bf \Pi}$ and
${\bf B}$ are measurable in the same units, thereby completing the
unification of electricity and magnetism that was begun by Maxwell.
The reduction to a single pair of fields and a single field-strength
tensor is an exquisite simplification of continuum electrodynamics.
\par
The changes to the theoretical treatment of electromagnetic fields
in linear media that have been presented here are stunning and
it is customary to propose experiments that would validate unordinary
theoretical results.
However, the existing experimental record related to the
Abraham--Minkowski controversy is an indication of serious
technical and conceptual difficulties in the measurement of continuum
electrodynamic phenomena related to macroscopic fields inside
materials, fields that cannot be measured directly.
On the other hand, the renormalization of time $t^{\prime}=t/n$ in a
linear medium is relatively easy to measure.
An open optical cavity, or free-space stolon, is a clock that ticks
once for every round-trip of a light pulse and that ticks more slowly
when it is immersed in a dielectric fluid.
\par
\section{Conclusion}
\par
The Maxwell equations are widely recognized as one of the most
successful theoretical treatments of our physical world, ever.
Then, the persistent inability to adequately address an important
issue in a simple continuous linear medium using the macroscopic Maxwell
theory should be seriously disturbing.
Instead the Abraham--Minkowski controversy has been treated as a
tautological curiosity in which some momentum quantity is
supplemented by another momentum quantity, as long as the
correct total momentum is achieved \cite{BIPfei,BIBarL}.
In the context of the open systems typically used in the treatment of
continuum electrodynamics, the Abraham--Minkowski controversy,
which is well-posed as a conservation issue, becomes ill-posed and
difficult to treat rigorously \cite{BIKran}.
The resolution of the Abraham--Minkowski dilemma in terms of a
self-consistent four-dimensional treatment of continuum electrodynamics
in a thermodynamically closed system based on the energy and momentum
conservation properties of the total energy--momentum tensor forces us
to confront a contradiction in the Maxwell formulation of classical
continuum electrodynamics.
This contradiction is not a minor issue as it resides at the very heart
of the relations between energy and momentum.
The contradiction is resolved by constructing a unified version of
continuum electrodynamics that is based on establishing consistency
between the equations of motion for macroscopic fields, the continuity
equations, the energy--momentum tensor, and the field-strength tensor
for a thermodynamically closed system with complete equations of
motion.
Like Newtonian dynamics is a low-velocity limit of Einstein relativity,
Maxwellian macroscopic electrodynamics must be realizable in some
limit of the more complete theory.
Indeed that is the case.
Natural materials are predominately dielectric or predominately
magnetic.
Fields inside materials cannot be measured directly and the two theories
yield identical observable results for a given refractive index $n$ for 
materials that can be characterized as either dielectric or magnetic,
but not as a combination of both.
\par

\end{document}